\font\twelverm=cmr10 scaled 1200    \font\twelvei=cmmi10 scaled 1200
\font\twelvesy=cmsy10 scaled 1200   \font\twelveex=cmex10 scaled 1200
\font\twelvebf=cmbx10 scaled 1200   \font\twelvesl=cmsl10 scaled 1200
\font\twelvett=cmtt10 scaled 1200   \font\twelveit=cmti10 scaled 1200
\skewchar\twelvei='177   \skewchar\twelvesy='60
\def\twelvepoint{\normalbaselineskip=12.4pt
  \abovedisplayskip 12.4pt plus 3pt minus 9pt
  \belowdisplayskip 12.4pt plus 3pt minus 9pt
  \abovedisplayshortskip 0pt plus 3pt
  \belowdisplayshortskip 7.2pt plus 3pt minus 4pt
  \smallskipamount=3.6pt plus1.2pt minus1.2pt
  \medskipamount=7.2pt plus2.4pt minus2.4pt
  \bigskipamount=14.4pt plus4.8pt minus4.8pt
  \def\rm{\fam0\twelverm}          \def\it{\fam\itfam\twelveit}%
  \def\sl{\fam\slfam\twelvesl}     \def\bf{\fam\bffam\twelvebf}%
  \def\mit{\fam 1}                 \def\cal{\fam 2}%
  \def\tt{\twelvett}
  \textfont0=\twelverm   \scriptfont0=\tenrm   \scriptscriptfont0=\sevenrm
  \textfont1=\twelvei    \scriptfont1=\teni    \scriptscriptfont1=\seveni
  \textfont2=\twelvesy   \scriptfont2=\tensy   \scriptscriptfont2=\sevensy
  \textfont3=\twelveex   \scriptfont3=\twelveex  \scriptscriptfont3=\twelveex
  \textfont\itfam=\twelveit
  \textfont\slfam=\twelvesl
  \textfont\bffam=\twelvebf \scriptfont\bffam=\tenbf
  \scriptscriptfont\bffam=\sevenbf
  \normalbaselines\rm}

\def\beginlinemode{\endmode
  \begingroup\parskip=0pt \obeylines\def\\{\par}\def\endmode{\par\endgroup}}
\def\beginparmode{\endmode
  \begingroup \def\endmode{\par\endgroup}}
\let\endmode=\par
{\obeylines\gdef\
{}}
\def\singlespace{\baselineskip=\normalbaselineskip}

\def\oneandtwofifthsspace{\baselineskip=\normalbaselineskip
  \multiply\baselineskip by 7 \divide\baselineskip by 5}
\def\oneandahalfspace{\baselineskip=\normalbaselineskip
  \multiply\baselineskip by 3 \divide\baselineskip by 2}
\def\doublespace{\baselineskip=\normalbaselineskip \multiply\baselineskip by 2}
\newcount\firstpageno
\firstpageno=2
\footline=
{\ifnum\pageno<\firstpageno{\hfil}\else{\hfil\twelverm\folio\hfil}\fi}
\let\rawfootnote=\footnote              
\def\footnote#1#2{{\rm\singlespace\parindent=0pt\rawfootnote{#1}{#2}}}
\def\raggedcenter{\leftskip=2em plus 12em \rightskip=\leftskip
  \parindent=0pt \parfillskip=0pt \spaceskip=.3333em \xspaceskip=.5em
  \pretolerance=9999 \tolerance=9999
  \hyphenpenalty=9999 \exhyphenpenalty=9999 }
\parskip=\medskipamount
\twelvepoint            
\overfullrule=0pt       
\def\preprintno#1{
 \rightline{\rm #1}}    
\def\author                     
  {\vskip 3pt plus 0.2fill \beginlinemode
   \singlespace \raggedcenter \twelvesc}
\def\affil                      
  {\vskip 3pt plus 0.1fill \beginlinemode
   \oneandahalfspace \raggedcenter \sl}
\def\abstract                   
  {\vskip 3pt plus 0.3fill \beginparmode
   \doublespace \narrower \noindent ABSTRACT: }
\def\endtitlepage               
  {\endpage                     
   \body}
\def\body                       
  {\beginparmode}               

\def\subhead#1{                 
  \vskip 0.25truein             
  {\raggedcenter #1 \par}
   \nobreak\vskip 0.1truein\nobreak}
\def\refto#1{$|{#1}$}           
\def\references                 
  {\subhead{References}         
   \beginparmode
   \frenchspacing \parindent=0pt \leftskip=1truecm
   \parskip=8pt plus 3pt \everypar{\hangindent=\parindent}}
\gdef\refis#1{\indent\hbox to 0pt{\hss#1.~}}    
\gdef\journal#1, #2, #3, 1#4#5#6{               
    {\sl #1~}{\bf #2}, #3, (1#4#5#6)}           
\def\refstylenp{                
  \gdef\refto##1{ [##1]}                                
  \gdef\refis##1{\indent\hbox to 0pt{\hss##1)~}}        
  \gdef\journal##1, ##2, ##3, ##4 {                     
     {\sl ##1~}{\bf ##2~}(##3) ##4 }}
\def\refstyleprnp{              
  \gdef\refto##1{ [##1]}                                
  \gdef\refis##1{\indent\hbox to 0pt{\hss##1)~}}        
  \gdef\journal##1, ##2, ##3, 1##4##5##6{               
    {\sl ##1~}{\bf ##2~}(1##4##5##6) ##3}}
\def\pr{\journal Phys. Rev., }

\def\prl{\journal Phys. Rev. Lett., }

\def\np{\journal Nucl. Phys., }
\def\npb{\journal Nucl. Phys., }
\def\pl{\journal Phys. Lett., }

\def\endreferences{\body}
\def\endpage                    
  {\vfill\eject}
\def\endpaper                   
  {\endmode\vfill\supereject}
\def\endit
  {\endpaper\end}
\def\ref#1{Ref. #1}                     
\def\Ref#1{Ref. #1}                     

\def\m@th{\mathsurround=0pt }
\font\twelvesc=cmcsc10 scaled 1200
\def\cite#1{{#1}}
\def\(#1){(\call{#1})}
\def\call#1{{#1}}
\def\taghead#1{}
\def\leaderfill{\leaders\hbox to 1em{\hss.\hss}\hfill}
\def\twiddle{\lower.9ex\rlap{$\kern-.1em\scriptstyle\sim$}}
\def\bigtwiddle{\lower1.ex\rlap{$\sim$}}
\def\gtwid{\mathrel{\raise.3ex\hbox{$>$\kern-.75em\lower1ex\hbox{$\sim$}}}}
\def\ltwid{\mathrel{\raise.3ex\hbox{$<$\kern-.75em\lower1ex\hbox{$\sim$}}}}
\def\square{\kern1pt\vbox{\hrule height 1.2pt\hbox{\vrule width 1.2pt\hskip 3pt
   \vbox{\vskip 6pt}\hskip 3pt\vrule width 0.6pt}\hrule height 0.6pt}\kern1pt}
\catcode`@=11
\newcount\tagnumber\tagnumber=0

\immediate\newwrite\eqnfile
\newif\if@qnfile\@qnfilefalse
\def\write@qn#1{}
\def\writenew@qn#1{}
\def\w@rnwrite#1{\write@qn{#1}\message{#1}}
\def\@rrwrite#1{\write@qn{#1}\errmessage{#1}}

\def\taghead#1{\gdef\t@ghead{#1}\global\tagnumber=0}
\def\t@ghead{}

\expandafter\def\csname @qnnum-3\endcsname
  {{\t@ghead\advance\tagnumber by -3\relax\number\tagnumber}}
\expandafter\def\csname @qnnum-2\endcsname
  {{\t@ghead\advance\tagnumber by -2\relax\number\tagnumber}}
\expandafter\def\csname @qnnum-1\endcsname
  {{\t@ghead\advance\tagnumber by -1\relax\number\tagnumber}}
\expandafter\def\csname @qnnum0\endcsname
  {\t@ghead\number\tagnumber}
\expandafter\def\csname @qnnum+1\endcsname
  {{\t@ghead\advance\tagnumber by 1\relax\number\tagnumber}}
\expandafter\def\csname @qnnum+2\endcsname
  {{\t@ghead\advance\tagnumber by 2\relax\number\tagnumber}}
\expandafter\def\csname @qnnum+3\endcsname
  {{\t@ghead\advance\tagnumber by 3\relax\number\tagnumber}}

\def\equationfile{%
  \@qnfiletrue\immediate\openout\eqnfile=\jobname.eqn%
  \def\write@qn##1{\if@qnfile\immediate\write\eqnfile{##1}\fi}
  \def\writenew@qn##1{\if@qnfile\immediate\write\eqnfile
    {\noexpand\tag{##1} = (\t@ghead\number\tagnumber)}\fi}
}

\def\callall#1{\xdef#1##1{#1{\noexpand\call{##1}}}}
\def\call#1{\each@rg\callr@nge{#1}}

\def\each@rg#1#2{{\let\thecsname=#1\expandafter\first@rg#2,\end,}}
\def\first@rg#1,{\thecsname{#1}\apply@rg}
\def\apply@rg#1,{\ifx\end#1\let\next=\relax%
\else,\thecsname{#1}\let\next=\apply@rg\fi\next}

\def\callr@nge#1{\calldor@nge#1-\end-}
\def\callr@ngeat#1\end-{#1}
\def\calldor@nge#1-#2-{\ifx\end#2\@qneatspace#1 %
  \else\calll@@p{#1}{#2}\callr@ngeat\fi}
\def\calll@@p#1#2{\ifnum#1>#2{\@rrwrite{Equation range #1-#2\space is bad.}
\errhelp{If you call a series of equations by the notation M-N, then M and
N must be integers, and N must be greater than or equal to M.}}\else%
 {\count0=#1\count1=#2\advance\count1
by1\relax\expandafter\@qncall\the\count0,%
  \loop\advance\count0 by1\relax%
    \ifnum\count0<\count1,\expandafter\@qncall\the\count0,%
  \repeat}\fi}

\def\@qneatspace#1#2 {\@qncall#1#2,}
\def\@qncall#1,{\ifunc@lled{#1}{\def\next{#1}\ifx\next\empty\else
  \w@rnwrite{Equation number \noexpand\(>>#1<<) has not been defined yet.}
  >>#1<<\fi}\else\csname @qnnum#1\endcsname\fi}

\let\eqnono=\eqno
\def\eqno(#1){\tag#1}
\def\tag#1$${\eqnono(\displayt@g#1 )$$}

\def\aligntag#1\endaligntag
  $${\gdef\tag##1\\{&(##1 )\cr}\eqalignno{#1\\}$$
  \gdef\tag##1$${\eqnono(\displayt@g##1 )$$}}

\def\eqalignno#1{\displ@y \tabskip\centering
  \halign to\displaywidth{\hfil$\displaystyle{##}$\tabskip\z@skip
    &$\displaystyle{{}##}$\hfil\tabskip\centering
    &\llap{$\displayt@gpar##$}\tabskip\z@skip\crcr
    #1\crcr}}

\def\displayt@gpar(#1){(\displayt@g#1 )}

\def\displayt@g#1 {\rm\ifunc@lled{#1}\global\advance\tagnumber by1
        {\def\next{#1}\ifx\next\empty\else\expandafter
        \xdef\csname @qnnum#1\endcsname{\t@ghead\number\tagnumber}\fi}%
  \writenew@qn{#1}\t@ghead\number\tagnumber\else
        {\edef\next{\t@ghead\number\tagnumber}%
        \expandafter\ifx\csname @qnnum#1\endcsname\next\else
        \w@rnwrite{Equation \noexpand\tag{#1} is a duplicate number.}\fi}%
  \csname @qnnum#1\endcsname\fi}

\def\ifunc@lled#1{\expandafter\ifx\csname @qnnum#1\endcsname\relax}

\let\@qnend=\end\gdef\end{\if@qnfile
\immediate\write16{Equation numbers written on []\jobname.EQN.}\fi\@qnend}

\catcode`@=12
\refstyleprnp
\catcode`@=11
\newcount\r@fcount \r@fcount=0
\def\refreset{\global\r@fcount=0}
\newcount\r@fcurr
\immediate\newwrite\reffile
\newif\ifr@ffile\r@ffilefalse
\def\w@rnwrite#1{\ifr@ffile\immediate\write\reffile{#1}\fi\message{#1}}

\def\writer@f#1>>{}
\def\referencefile{
  \r@ffiletrue\immediate\openout\reffile=\jobname.ref%
  \def\writer@f##1>>{\ifr@ffile\immediate\write\reffile%
    {\noexpand\refis{##1} = \csname r@fnum##1\endcsname = %
     \expandafter\expandafter\expandafter\strip@t\expandafter%
     \meaning\csname r@ftext\csname r@fnum##1\endcsname\endcsname}\fi}%
  \def\strip@t##1>>{}}

\def\citeall#1{\xdef#1##1{#1{\noexpand\cite{##1}}}}
\def\cite#1{\each@rg\citer@nge{#1}}     

\def\each@rg#1#2{{\let\thecsname=#1\expandafter\first@rg#2,\end,}}
\def\first@rg#1,{\thecsname{#1}\apply@rg}       
\def\apply@rg#1,{\ifx\end#1\let\next=\relax
\else,\thecsname{#1}\let\next=\apply@rg\fi\next}

\def\citer@nge#1{\citedor@nge#1-\end-}  
\def\citer@ngeat#1\end-{#1}
\def\citedor@nge#1-#2-{\ifx\end#2\r@featspace#1 
  \else\citel@@p{#1}{#2}\citer@ngeat\fi}        
\def\citel@@p#1#2{\ifnum#1>#2{\errmessage{Reference range #1-#2\space is bad.}%
    \errhelp{If you cite a series of references by the notation M-N, then M and
    N must be integers, and N must be greater than or equal to M.}}\else%
 {\count0=#1\count1=#2\advance\count1
by1\relax\expandafter\r@fcite\the\count0,%
  \loop\advance\count0 by1\relax
    \ifnum\count0<\count1,\expandafter\r@fcite\the\count0,%
  \repeat}\fi}

\def\r@featspace#1#2 {\r@fcite#1#2,}    
\def\r@fcite#1,{\ifuncit@d{#1}
    \newr@f{#1}%
    \expandafter\gdef\csname r@ftext\number\r@fcount\endcsname%
                     {\message{Reference #1 to be supplied.}%
                      \writer@f#1>>#1 to be supplied.\par}%
 \fi%
 \csname r@fnum#1\endcsname}
\def\ifuncit@d#1{\expandafter\ifx\csname r@fnum#1\endcsname\relax}%
\def\newr@f#1{\global\advance\r@fcount by1%
    \expandafter\xdef\csname r@fnum#1\endcsname{\number\r@fcount}}

\let\r@fis=\refis                       
\def\refis#1#2#3\par{\ifuncit@d{#1}
   \newr@f{#1}%
   \w@rnwrite{Reference #1=\number\r@fcount\space is not cited up to now.}\fi%
  \expandafter\gdef\csname r@ftext\csname r@fnum#1\endcsname\endcsname%
  {\writer@f#1>>#2#3\par}}

\def\ignoreuncited{
   \def\refis##1##2##3\par{\ifuncit@d{##1}%
   \else\expandafter\gdef\csname r@ftext\csname r@fnum##1\endcsname\endcsname%
     {\writer@f##1>>##2##3\par}\fi}}

\def\r@ferr{\endreferences\errmessage{I was expecting to see
\noexpand\endreferences before now;  I have inserted it here.}}
\let\r@ferences=\references
\def\references{\r@ferences\def\endmode{\r@ferr\par\endgroup}}

\let\endr@ferences=\endreferences
\def\endreferences{\r@fcurr=0
  {\loop\ifnum\r@fcurr<\r@fcount
    \advance\r@fcurr by 1\relax\expandafter\r@fis\expandafter{\number\r@fcurr}%
    \csname r@ftext\number\r@fcurr\endcsname%
  \repeat}\gdef\r@ferr{}\global\r@fcount=0\endr@ferences}

\let\r@fend=\endpaper\gdef\endpaper{\ifr@ffile
\immediate\write16{Cross References written on []\jobname.REF.}\fi\r@fend}

\catcode`@=12

\citeall\refto          
\citeall\ref            %
\citeall\Ref            %

\referencefile

\def\ssc{\scriptscriptstyle}
\def\gR{g_{\ssc R}}
\def\gB{g_{\ssc B-L}}
\def\gRB{g_{\ssc R,B-L}}
\def\gBR{g_{\ssc B-L,R}}
\def\bR{b_{\ssc R}}
\def\bB{b_{\ssc B-L}}
\def\bRB{b_{\ssc R,B-L}}
\def\half{{\textstyle{ 1\over 2}}}
\def\frac#1/#2{{\textstyle{#1 \over #2}}}

\def\umichadd{Randall Physics Laboratory\\University of Michigan
\\Ann Arbor MI 48109-1120}

\def\oneandtwofifthsspace{\baselineskip=\normalbaselineskip
  \multiply\baselineskip by 29 \divide\baselineskip by 20}

\font\titlefont=cmr10 scaled\magstep3
\def\bigtitle                      
  {\null\vskip 3pt plus 0.2fill
   \beginlinemode \doublespace \raggedcenter \titlefont}


\oneandahalfspace
\preprintno{hep-ph/9602349}
\bigtitle{Implications of supersymmetric models with
            natural R-parity conservation}
\bigskip

\author Stephen P.~Martin
\affil{\umichadd}

\body

\abstract \oneandahalfspace
In the minimal supersymmetric standard model, the conservation of
$R$-parity is phenomenologically desirable, but is {\it ad hoc}
in the sense that it is not required for the internal consistency of
the theory. However, if $B-L$ is gauged at very high energies,
$R$-parity will be conserved automatically and exactly, provided only that all
order parameters carry even integer values of $3(B-L)$.
We propose a minimal extension of the
supersymmetric standard model in which $R$-parity conservation arises
naturally in this way. This approach predicts the existence of a
very weakly coupled, neutral chiral supermultiplet of particles
with electroweak-scale
masses and lifetimes which may be cosmologically interesting.
Neutrino masses arise via an intermediate-scale seesaw mechanism,
and a solution to the $\mu$ problem is naturally incorporated.
The apparent unification of gauge couplings at high energies
is shown to be preserved in this approach. We also
discuss a next-to-minimal extension, which predicts
a pair of electroweak-scale chiral supermultiplets with electric charge 2.

\endtitlepage
\oneandahalfspace

\subhead{1. Introduction}
\taghead{1.}

One of the successes of the Standard Model of particle physics is the
automatic conservation of baryon number ($B$) and total lepton number ($L$)
at the renormalizable level. These conservation laws follow simply from the
particle content and $SU(3)_C \times SU(2)_L \times U(1)_Y$
gauge invariance, and do not entail additional assumptions.

The simplest supersymmetric extension of the Standard Model does
not share this appealing feature, because the existence of scalar partners
of the quarks and leptons allows for renormalizable violation of $B$ and
$L$. The most general renormalizable and
$SU(3)_C \times SU(2)_L \times U(1)_Y$-invariant superpotential is given
schematically by
$$
\eqalign{
& W = W_0 + W_1 + W_2\> ,\cr
& W_0 = \mu  H_u H_d + H_u Q u + H_d Q d + H_d L e\> ,\cr
& W_1 = udd \> ,\qquad\qquad  W_2 = \mu^\prime L H_u + QLd + LLe
\> .\cr}
$$
[Here $Q$ and $L$ are chiral superfields for the $SU(2)_L$-doublet quarks and
leptons; $u$, $d$, $e$ are chiral
superfields for the $SU(2)_L$-singlet quarks
and leptons, and $H_u$, $H_d$
are the two $SU(2)_L$-doublet Higgs chiral superfields.
It is possible to eliminate
$\mu^\prime$ by a suitable rotation among the superfields $H_d$ and $L$;
but in general it is somewhat misleading to do so
because in many extensions of
minimal supersymmetry $H_d$ and $L$ will not have the same quantum numbers, and
the appropriate rotation cannot be performed.] The terms in $W_0$ are just the
supersymmetric versions of the usual standard
model Yukawa couplings and Higgs mass, and they conserve
$B$ and $L$. However, $W_1$ violates $B$ by one unit and $W_2$
violates $L$ by one unit. To prevent the proton from decaying within
seconds or hours,
either the couplings in $W_1$ or those in $W_2$ (or both)
must be extremely small.
In this sense, the supersymmetric standard model appears to be less successful
or at least less elegant than the Standard Model, since the observed
conservation of $B$ and $L$ is no longer automatic, but requires some
additional assumptions about the structure of the theory.

The most common way to save the proton from the supersymmetric threat is to
forbid all of the terms occurring in $W_1$ and $W_2$ by imposing the
discrete $Z_2$ symmetry [\cite{rparity},\cite{rparity2}]
known as $R$-parity or matter parity. The matter
parity of each superfield may be defined as
$$
({\rm matter}\>\>{\rm parity}) \equiv (-1)^{3(B-L)} \> .
\eqno(matterparity)
$$
Then multiplicative conservation of matter parity forbids all terms
in $W_1$ and $W_2$, while allowing the phenomenologically necessary
ones in $W_0$. Equivalently, the $R$-parity of any component field is
defined by $(-1)^{3(B-L) + 2 s}$, where $s$ is the spin of the field.
Since $(-1)^{2s} $ is of course conserved in any Lorentz-invariant
interaction, matter parity conservation and $R$-parity conservation
are precisely equivalent. The description in terms of matter parity
makes clear that there is nothing intrinsically ``$R$-symmetric"
about this symmetry; in other words, it admits a formulation at the
superfield level. Conversely, the description in terms of $R$-parity is
convenient in phenomenological discussions, because it happens that all
Standard Model states have $R$-parity $+1$, while all superpartners have
$R$-parity $-1$. Conservation of $R$-parity then immediately implies
that superpartners can be produced only in pairs, and that the lightest
supersymmetric particle (LSP) is absolutely stable.
\phantom{\cite{alternative}\cite{baryonparity}\cite{CM}\cite{violation1}
\cite{violation2}\cite{violation3}}

The minimal supersymmetric standard model (MSSM) with $R$-parity
conservation
can provide a description of nature which is consistent with all known
observations. However, the assumption of $R$-parity conservation might
appear to be {\it ad hoc}, since it is not required for the internal
consistency of the theory. Alternative discrete symmetries have in fact
been proposed (see for example [\cite{alternative},\cite{baryonparity}]).
Perhaps the simplest of these
is the $Z_3$ discrete ``baryon parity"
of Ib\'a\~nez and Ross [\cite{baryonparity}], which turns out to imply the
falsifiable predictions that the proton is {\it absolutely}
stable and there can be no neutron--antineutron oscillations even if there are
isosinglet quark superfields near the TeV scale [\cite{CM}].
One might also entertain the possibility of
small $R$-parity violation, with intriguing phenomenological consequences
(see for example [\cite{violation1}-\cite{violation4}]).
However, if $R$-parity is not exact, the LSP is unstable and so cannot be
a candidate for the cold dark matter, unless its lifetime
is of order the age of the universe.
\phantom{\cite{Mohapatra},\cite{FIQ},\cite{discreteanomalies}}

Fortunately, there is a particularly compelling scenario which does
automatically provide for exact $R$-parity conservation due to a deeper
principle. This is suggested immediately by \(matterparity), which shows
that matter parity is simply a $Z_2$ subgroup of $B-L$. If $U(1)_{B-L}$
is gauged at high energies, it will forbid each of the terms in $W_1$ and
$W_2$ [\cite{Mohapatra}-\cite{sscfgrp}].
Of course, there is no massless gauge boson found in nature which
couples to $B-L$, so $U(1)_{B-L}$ must be spontaneously broken.
The question then becomes how to break $B-L$ without also breaking
matter parity. To guarantee that matter parity should remain unbroken
even after a gauged $U(1)_{B-L}$  is broken, it is necessary and
sufficient to require that all scalar vacuum expectation values (VEVs)
or other order parameters carry $3(B-L)$ charges which are even integers.
Following the general arguments of Krauss and Wilczek [\cite{KW}],
the gauged $U(1)_{B-L}$ symmetry breaks down to a $Z_2$ subgroup which,
in view of \(matterparity),
is nothing other than matter parity. Unlike a global discrete symmetry,
such a gauged discrete symmetry must be respected by Planck scale effects, and
satisfies discrete anomaly cancellation conditions
[\cite{discreteanomalies},\cite{BD},\cite{more}].
Note that it is a contradiction in terms to speak of explicit
$R$-parity breaking in a supersymmetric model with gauged $U(1)_{B-L}$;
$R$-parity will either be exactly conserved
(if all order parameters carry
only even integer values of $3(B-L)$) or spontaneously broken
[\cite{spon}] (if some
order parameter carries an odd integer $3(B-L)$).

Of course, this scenario for the origin of matter parity is hardly
mandatory, since it is technically natural to forbid the terms in $W_1$ and
$W_2$ ``by hand" as an unexplained assumption. However, it is worthwhile
to take seriously the idea that $R$-parity conservation is explainable,
since we then obtain some quite non-trivial information about physics at
very high energy scales. Not only do we gain an indication that the
unbroken gauge group should contain $U(1)_{B-L}$, but we also obtain
information about how it should (and should not!) be broken.

In ref.~[\cite{sscfgrp}],
the criteria for maintaining natural $R$-parity conservation in
models with gauged $U(1)_{B-L}$ were considered for various extended gauge
groups. Consider,
for example, the possibility of a natural explanation for $R$-parity
conservation in a supersymmetric $SO(10)$ grand unified theory (GUT). Now,
$SO(10)$ contains $U(1)_{B-L}$ as a subgroup, so that $R$-parity conservation
is automatic before spontaneous symmetry breaking. However, the smallest
``safe" representations for a scalar which can break $U(1)_{B-L}$ without
breaking the gauged matter parity subgroup in the process are
$\bf 120,126,210\ldots $
and their conjugates. This is unfortunate, since experience has shown
that it is quite difficult to
build a successful supersymmetric GUT with such large representations.
The alternative is to break
$SO(10)$ with an order parameter in a $\bf 16$ representation, and that is
what is usually done. However, the neutral component of a $\bf 16$ carries
$3(B-L) = 3$, so that the original automatic $R$-parity conservation is
forfeit. Indeed, renormalizable matter parity violation appears in the
low energy superpotential from non-renormalizable operators of the form
$(1/M) \langle {\bf 16} \rangle \times {\bf 16}
\times {\bf 16} \times {\bf 16}$.
(Here and in the following $M$ is some physical cutoff scale,
perhaps $M = M_{\rm Planck}/\sqrt{8 \pi}$.)
As another example, one might consider
an extension of the MSSM with a Pati-Salam gauge group
$SU(4)_{PS} \times SU(2)_L \times U(1)_R$. Again, $R$-parity conservation
is automatic before spontaneous symmetry breaking since
$SU(4)_{PS} \supset SU(3)_C \times U(1)_{B-L}$. To avoid breaking
matter parity in the process of breaking $SU(4)_{PS}$, it is necessary
and sufficient that all order parameters have even $SU(4)_{PS}$ quadrality,
since $SU(4)_{PS}$ quadrality $= 3(B-L)$ [mod 4]. The smallest such ``safe"
representation for an order parameter which breaks $U(1)_{B-L}$
is the $\bf 10$ of $SU(4)_{PS}$.

In ref.~[\cite{KM2}], the dynamical issues associated with
automatic $R$-parity conservation have been considered in the
case of left-right symmetric models.
It was found that in a wide class of such models, $R$-parity
must be spontaneously broken because of the form of the scalar
potential, although this can be evaded if non-renormalizable
interactions are included.

In this paper, we will consider instead a minimal extension of the MSSM
in which the gauge group is extended by only $U(1)_{B-L}$.
Anomaly cancellation for $U(1)_{B-L}$ implies the existence of three
neutrino chiral superfields $\nu$ which carry $B-L = 1$ and
are singlets of the standard model gauge group.
A VEV for the scalar component of $\nu$ would spontaneously break matter
parity, so we will require it to be absent.
While such a weak-scale VEV for $\nu$ is not yet ruled out phenomenologically,
we adopt for this paper
the point of view that this is unacceptable, since we want to
explore here only possibilities with exact and automatic $R$-parity
conservation.

To obtain a realistic theory of
neutrino masses, we may invoke the seesaw mechanism
[\cite{seesaw}] by means
of the superpotential
$$
W \supset y_\nu H_u L \nu + {y_S\over 2} S  \nu \nu \> .
\eqno(seesaw)
$$
Here $S$ is a chiral superfield which must carry $B-L = -2$. Assuming
that $\langle S \rangle $ is much larger than the electroweak scale
$m_W$,
one finds that
the lighter neutrino mass eigenstates have tiny masses
$\sim (y_\nu \langle H_u\rangle)^2/(y_S \langle S \rangle)$.
Now, the role of $S$ within this framework might
be played by a composite field
$S = {\overline\nu }\, {\overline\nu} /M$. However, this again
cannot be consistent with our criteria for automatic $R$-parity conservation,
since then a VEV $\langle S \rangle \not= 0$
implies a VEV $\langle {\overline \nu} \rangle \not= 0$.
Therefore, we prefer the possibility
that $S$ is a fundamental chiral superfield, so that
the VEV $\langle S \rangle$ cannot break the matter parity subgroup
of $U(1)_{B-L}$. The field $S$ must
be accompanied by a field $\overline S$ in the conjugate
representation, in order to cancel the anomalies and to allow spontaneous
symmetry breaking in a nearly $D$-flat direction. (Otherwise there would
be catastrophically large supersymmetry-breaking $D$-terms, which would
destabilize the electroweak scale.)
For the models in this paper, the scale
$\langle S \rangle$ is an intermediate one, roughly the geometric
mean between the electroweak scale and the Planck scale.
Assuming that the Yukawa couplings $y_\nu$ are of the same
order as those of the charged leptons,  one then expects
light neutrino masses in
the range relevant [\cite{Langacker}] to solar or
atmospheric neutrino oscillations and hot dark matter.

In section 2 of
this paper we will propose a minimal extension of the
supersymmetric standard model which
successfully implements automatic and unbroken $R$-parity conservation
from gauged $B-L$, and discuss some of its implications. Section 3 contains
some discussion of the subtleties associated with $U(1)$ mixing in this model,
and the effect of intermediate scale thresholds on the sparticle spectrum.
In section 4 we will discuss a  next-to-minimal model [with gauge group
$SU(3)_C \times SU(2)_L \times SU(2)_R \times U(1)_{B-L}$].
Section 5 contains some concluding remarks.

\subhead{2. A minimal model of automatic $R$-parity conservation}
\taghead{2.}

We consider a supersymmetric model with gauge group
$SU(3)_C\times SU(2)_L \times U(1)_Y \times U(1)_{B-L}$. The
MSSM chiral superfields (plus $\nu$) transform under this gauge group
as three copies of
$$
\eqalign{
Q &\sim ({\bf 3}, {\bf 2}, \frac1/6, \frac1/3 ) \cr
L &\sim ({\bf 1}, {\bf 2}, -\frac1/2, -1 ) \cr
}\qquad\>\>\>
\eqalign{
d &\sim (\overline{\bf 3}, {\bf 1}, \frac1/3, -\frac1/3 ) \cr
e &\sim ({\bf 1}, {\bf 1}, 1, 1 ) \cr
}\qquad\>\>\>
\eqalign{
u &\sim (\overline{\bf 3}, {\bf 1}, -\frac2/3, -\frac1/3 ) \cr
\nu &\sim ({\bf 1}, {\bf 1}, 0, 1 )\cr
}
$$
and two Higgs doublets
$$
H_u \sim ({\bf 1}, {\bf 2}, \frac1/2, 0 )
\qquad\qquad
H_d \sim ({\bf 1}, {\bf 2}, -\frac1/2, 0 )\> .
$$
In order to break $U(1)_{B-L}$, we introduce two chiral superfields
$$
S \sim ({\bf 1}, {\bf 1}, 0, -2 )
\qquad\qquad
{\overline S} \sim ({\bf 1}, {\bf 1}, 0, 2 )\> .
$$
Because this gauge group contains two abelian factors, in general
one must consider the possibility of arbitrary mixing between
$U(1)_Y$ and $U(1)_{B-L}$. (Indeed, it would be quite unexpected if the
gauge interactions were diagonal in the $Y$, $B-L$ basis.)
We will be able to postpone a discussion of this until the next section,
however.

Besides the interaction between $S$ and $\nu$ given in \(seesaw),
$S$ and $\overline S$ participate in a non-renormalizable superpotential
interaction\footnote{$^\dagger$}{We do not allow tree-level
superpotential mass terms $S\overline S$ and $H_u H_d$ or their soft
supersymmetry breaking counterparts,
in sympathy with general results in superstring models.}
$$
W =  {\lambda \over 2 M} S^2 {\overline S}^2
\eqno(superpotential)
$$
and soft supersymmetry-breaking terms
$$
V_{\rm soft} = m^2 |S|^2 + {\overline m}^2 | \overline S |^2 -
\left ( {A \over 2 M} S^2 {\overline S}^2 + c.c. \right ) \> .
\eqno(vsoft)
$$
(We use the same symbol for each chiral superfield and its scalar component.)
The parameters $m^2$, ${\overline m}^2$ and $A$ should each be of order the
electroweak scale $m_W$ in order not to upset the hierarchy.
Note that the $A/M$ term, while dimensionless, should nevertheless
be treated as ``soft" because of its tiny magnitude. Such terms should
naturally arise in supergravity models, and this one will play a crucial
role on several accounts, as we shall soon see. By a suitable phase rotation,
we take $A$ to be real and positive, while the phase of $\lambda$ can be
arbitrary.
The full scalar potential for the $S$ and $\overline S$
degrees of freedom is given by the sum of $V_{\rm soft}$ and
$$
V_{\rm SUSY} = {|\lambda |^2 \over M^2} |S{\overline S}|^2
( |S|^2 +
|{\overline S} |^2 ) + {g_{X}^2  \over 2} \left (
|S|^2 - |{\overline S}|^2 \right )^2
\eqno(vsusy)
$$
where the latter term is the $D$-term contribution. The parameter $g_X$
is related to the gauge couplings of the $U(1)$s,
and will be explicitly identified in the next section when
we discuss the effects of $U(1)$ mixing.
\phantom{\cite{MSY}\cite{Drees}\cite{HK}\cite{FHKN}\cite{KT}}

A familiar method of inducing spontaneous gauge
symmetry breaking in supersymmetric models is
to arrange for the
running soft (mass)$^2$ of the appropriate scalar to become negative
at some scale. In the usual MSSM, this radiative symmetry breaking is achieved
by means of a large top-quark Yukawa coupling which drives the Higgs
(mass)$^2$ negative. In the model discussed here,
the parameter $m^2$ obtains
a negative radiative correction due to the Yukawa coupling $y_S$ in
\(seesaw), and this has been exploited to obtain radiative symmetry breaking in
similar models [\cite{MSY}-\cite{KM}].
However, in the present case
one also obtains a large {\it positive} radiative
correction to both $m^2$ and ${\overline m}^2$ from $U(1)$-gaugino
loops.
Indeed, the large ($\pm 2$) $B-L$ charges of $S$ and $\overline S$
make it seem somewhat problematic to achieve radiative symmetry
breaking in the traditional way; an examination of the renormalization
group (RG) equations shows that very large (or numerous)
Yukawa couplings $y_S$ seem to be required.
Fortunately, it is not really necessary
for $m^2$ or ${\overline m}^2$ to be driven negative in this model,
since the $A$ term in \(vsoft) always
favors spontaneous symmetry breaking.
A non-trivial local minimum of the scalar potential will be obtained
provided that
$$
A^2 - 6 |\lambda |^2 (m^2 + {\overline m}^2 ) > 0 \> .
\eqno(stability0)
$$
This minimum will be global if
$$
A^2 - 8|\lambda |^2 (m^2 + {\overline m}^2 ) > 0 \> .
\eqno(stabilityglobal)
$$
These conditions can be satisfied either by driving $m^2$ negative, or
simply by taking the free parameter $A$ to be sufficiently large
(while still roughly of order the electroweak scale),
or perhaps by a combination of these effects. In any case,
the minimum of the potential occurs
along a nearly $D$-flat direction:
$$
\langle S \rangle^2 \approx
\langle {\overline S} \rangle^2 \approx
{M \over 6 |\lambda |^2} \left ( A + \sqrt{ A^2 - 6 |\lambda |^2 (m^2 +
{\overline m}^2 ) } \right )
\eqno(vevS)
$$
with the deviation from $D$-flatness given by
$$
\langle S \rangle^2 -
\langle {\overline S} \rangle^2 \approx
({\overline m}^2 - m^2)/(2 g_{X}^2) \> .
\eqno(diffvev)
$$
We see from \(vevS) that the characteristic scale of $B-L$ breaking
is roughly a
geometric mean between the electroweak scale and the Planck scale:
$(m_W M)^{1/2} \sim 10^{10}$ GeV.
Since only $S$ and $\overline S$ obtain VEVs, it is clear that
$R$-parity conservation is automatic (and in fact unavoidable) in this
model. Note that the minimum of the scalar potential is stable against
$\nu$ obtaining a VEV (and more generally against arbitrary perturbations
of $S$, $\overline S$ and $\nu$);
this can be understood from the fact that the scalar potential
contains large positive
semi-definite contributions $|F_\nu|^2 = |y_S \langle S \rangle \nu  |^2$.

After spontaneous symmetry breaking, a
gauge boson and gaugino obtain masses
$$
M_{I} = 2 g_X \langle S \rangle \eqno(mBL)
$$
by eating the would-be Nambu-Goldstone boson degree of freedom
$ {\rm Im} [ S  - {\overline S} ]/\sqrt{2}$.
(Without loss of generality, we take $\langle S \rangle$,
$\langle {\overline S} \rangle $
to be real and positive.) Here $g_X$ is the same coupling appearing
in \(vsusy).
In addition, one real scalar degree of freedom
(given approximately by
$ {\rm Re} [ (S - \langle S \rangle)  - ({\overline S} - \langle {\overline S}
\rangle ) ]/\sqrt{2}$) and one Weyl fermion
from $S,\overline S$ get masses $M_{I}$, forming a complete massive
vector supermultiplet. There remains one light neutral chiral
supermultiplet, given approximately by
$$ \Phi \approx
[ S - \langle S \rangle  + {\overline S} - \langle {\overline S}
\rangle  ]/\sqrt{2},
\eqno(defphi)
$$
whose components all obtain electroweak-scale masses. These degrees of freedom
consist of a Weyl fermion $\psi$ (with $R$-parity $-1$) of mass
$$
m_\psi = 3 |\lambda | {\langle S \rangle^2 \over M}
\eqno(psimass)
$$
and two real scalar degrees of freedom $a$ and $b$ (of $R$-parity +1)
with squared masses
$$
m_a^2 = 4 A {\langle S \rangle^2 \over M }, \qquad\>\>\>\>\>
m_b^2 =
2 A {\langle S \rangle^2 \over M } - 2(m^2 + {\overline m}^2)\> .
\eqno(massab)
$$

Let us pause to remark on several interesting features of this spectrum
of electroweak-scale, neutral particles. First, note that in the limit
$A \rightarrow 0$, $m_a^2$ vanishes and the scalar $a$ becomes
a Nambu-Goldstone boson. This corresponds to the spontaneous
breaking of a continuous $R$-symmetry
of the superpotential, which is explicitly broken only by the $A$ term.
Fortunately, there is no reason for the parameter $A$ to be small compared
to the electroweak scale; on the contrary, it is likely that $A$ should be
large in order to achieve the necessary condition
\(stability0) for spontaneous symmetry breaking, as we have already
discussed. Perhaps a more plausible limit physically is
$A \gg |\lambda|^2 (m^2 + {\overline m}^2)$ (but still very roughly
of order the electroweak scale), which leads to
$$
\langle S \rangle^2 \approx {A M\over 3 |\lambda |^2}; \qquad
m_\psi \approx {A\over |\lambda |};
\qquad m_a \approx \sqrt{4\over 3}\, {A\over |\lambda |};
\qquad m_b \approx \sqrt{2\over 3}\, {A\over |\lambda |}.
\eqno(largeA)
$$
In general, the masses of the
component fields of the supermultiplet $\Phi$ satisfy the sum rule
$$
m_a^2 + m_b^2 - 2 m_\psi^2 = m^2 + {\overline m}^2 \> .
\eqno(phisumrule)
$$
It is not difficult to show that the lightest member of the supermultiplet
$\Phi$ is always one of the scalars ($a$ or $b$).

An important byproduct of this symmetry breaking scenario follows from the
existence of an allowed term in the non-renormalizable superpotential
which is of the same order as \(superpotential):
$$
W \supset {\lambda^\prime\over M} H_u H_d S \overline S \> .
\eqno(urmu)
$$
After symmetry breaking, one obtains the usual $\mu H_u H_d$ term of the MSSM,
with
$$
\mu = \lambda^\prime \langle S \rangle^2/M \eqno(mu)
$$
which is naturally of order the electroweak scale. The corresponding
soft MSSM Higgs mass term (often denoted $B\mu$) is generated in
the same way from the soft term corresponding to \(urmu).
This is a solution to the problem of generating an electroweak-scale
$\mu$ term along the lines of [\cite{KN}].

The interaction
\(urmu) also plays another crucial role in this model; it allows
$a$, $b$, and $\psi$ to decay in a
cosmologically timely fashion. These fields clearly
have only tiny couplings
to the particles of the MSSM. The most important such interactions
actually follow from \(urmu); one finds the
coupling of $\Phi \supset (a,b,\psi )$ to MSSM states
$$
W \supset { \sqrt{2} \lambda^\prime\>\langle S \rangle \over M}
\Phi H_u H_d \> .
\eqno(phihuhd)
$$
The dimensionless coupling
$\sqrt{2}\lambda^\prime \>\langle S \rangle / M$ is very roughly
of order $(m_W/M)^{1/2}$ if $\lambda^\prime$ is of order unity,
leading to decay widths of order
$\Gamma \sim m_W^2/(8 \pi M) \sim 10^{-15}$ GeV for
the component fields of
$\Phi$ into MSSM states.
Depending on kinematic and mixing angle factors,
one has\footnote{$^\dagger$}{The components of $\Phi$ can also decay into light
(s)neutrino pairs, but these decays turn out not to be
competitive because they
are suppressed by the seesaw mixing angle squared.}
two-body decays
$a,b \rightarrow$ Higgs + Higgs or
$a,b \rightarrow$ Higgsino + Higgsino and
$\psi \rightarrow $ Higgs + Higgsino
decays with lifetimes of order $10^{-9}$
seconds, give or take several orders
of magnitude. (This should be compared to the notorious problem of
the lifetime of an electroweak-scale gravitino,
which can be estimated to be of order $M/m_W$ times as long.)
Thus it is unlikely that late decays of these particles could jeopardize
the successful predictions of big-bang nucleosynthesis, although they might
certainly have other interesting cosmological effects which should
be carefully investigated. Note in particular
that each decay of $\psi$ results in the production of one stable
LSP. (It does not appear viable to allow $\psi$ itself to be
the LSP, because its annihilation cross-section is so tiny that it
would cause the universe to become matter dominated too early.)

\subhead{3. U(1) mixing and the sparticle spectrum}
\taghead{3.}

The model described in the previous section contains two $U(1)$ factors
which can mix in an {\it a priori} arbitrary way. We chose to specify
the charges of the chiral superfields in the $Y$, $B-L$ basis, but this
does not completely specify the gauge interactions of these fields.
In fact, it would be rather surprising if the gauge interactions at high
energies were diagonal in this basis. We will choose instead to use
the ``$SO(10)$-inspired" basis given by
$U(1)_R$, $U(1)_{B-L}$, with the $B-L$ charges as before,
and $R= Y-(B-L)/2$. [There would be no mixing in this basis if the gauge group
were imbedded in e.g.~unbroken $SO(10)$.]
One can always perform a rotation
on the $U(1)$ vector supermultiplets so that the kinetic terms are
diagonal and canonically normalized for the $U(1)_R$, $U(1)_{B-L}$
gauge bosons and gauginos.
Then the interactions with matter fields $\phi_i$ are
specified by the covariant derivative
$$
\eqalign{
 D_\mu \phi_i &= (\partial_\mu + i{\overline g}_i^{\ssc R} A^{\ssc R}_\mu
+ i {\overline g}_i^{\ssc B-L} A^{\ssc B-L}_\mu ) \phi_i \; ,\cr
{\overline g}_i^{\ssc R} &= \gR R_i+ \gBR \sqrt{3/8}\; (B-L)_i \; ,\cr
{\overline g}_i^{\ssc B-L} &=  \gB \sqrt{3/8}\; (B-L)_i + \gRB R_i\> .
\cr }
$$
The charges $R_i$ and $(B-L)_i$ are constants and are not renormalized.
However, in general the couplings $\gB$, $\gR$, $\gBR$, and $\gRB$
all require counterterms and are renormalized [\cite{mixing}].
The mixing couplings
$\gBR$ and $\gRB$ cannot avoid counterterms unless the matter content
is special, e.g. in complete multiplets of a non-abelian group containing
at least one of the $U(1)$'s.
It is therefore not consistent in general, and in particular in the model
of the previous section, to set $\gBR$ and $\gRB$
equal to 0. At any particular renormalization scale one can perform a rotation
on the vector superfield basis to set either $\gBR$ {\it or} $\gRB$
equal to 0 [\cite{mixing}].
This condition is not renormalization scale-invariant, however,
so it is sometimes
convenient to keep all four couplings as free parameters.

In terms of these parameters, the coupling $g_X$ appearing in the previous
section is
$$
g_X^2 = (\gR - \sqrt{3/ 2}\; \gBR)^2 + (\sqrt{3/2}\; \gB - \gRB)^2\> .
\eqno(gXdef)
$$
At the scale of symmetry breaking, the surviving $U(1)_Y$ gauge coupling
is given by (in a GUT-like normalization)
$$
g_Y = \sqrt{5/2}\; (\gR\gB - \gBR\gRB )/g_X\> .\eqno(defgy)
$$
The one-loop RG equations for the gauge couplings are
[$t={\rm ln}(Q/Q_0)$]:
$$
{d\over dt} \pmatrix{\gB & \gBR\cr\gRB & \gR} =
{1\over 16 \pi^2}
\pmatrix{\gB & \gBR\cr\gRB & \gR} \pmatrix{\bB&\bRB\cr\bRB & \bR}\> ,
\eqno(matrixrges)
$$
with
$$
\eqalign{
&\bB = 9 (\gB^2 + \gRB^2) - 2 \sqrt{6}\,\gB \gRB \> ,
\qquad\>\>\>\>
\bR = 9(\gR^2 + \gBR^2) - 2 \sqrt{6}\, \gR \gBR \> ,\cr
&\bRB = 9 (\gR \gRB + \gB \gBR) - \sqrt{6}\,
(\gB \gR+ \gBR \gRB ) \> .\cr}
$$
Using these equations, one finds that $g_Y$ defined
by \(defgy) satisfies the one-loop RG equation
$$
{d \over dt} g_Y = {1\over 16 \pi^2}{33\over 5} g_Y^3
\eqno(Yrges)
$$
(just as in the MSSM) both above and below $M_I$,
so that the condition for unification of $g_Y$ with the
$SU(3)_C$ and $SU(2)_L$ gauge couplings $g_3$ and $g_2$ is unaffected
by mixing, up to two-loop and threshold effects.

It therefore is sensible to impose a gauge coupling unification condition
on all the couplings, as could follow from a superstring or a GUT model.
At the unification scale $t_U$, one might
therefore take
$$
g_3 = g_2 = \gR = \gB \equiv g_{\ssc U},\qquad\>\>\> \gBR = \gRB = 0\> .
\eqno(bcs)
$$
We will assume these boundary conditions for the remainder of this
section, although it cannot be overemphasized that alternative
boundary conditions are certainly possible.
At lower scales, one can then solve the one-loop RG equations analytically
(for example by rotating to the multiplicatively renormalized basis),
with the result
$$
\eqalignno{
&\gR = \gB = g_{\ssc U}\;(\kappa_+ + \kappa_- )/2\> , &(grsol) \cr
&\gRB = \gBR = g_{\ssc U}\;(\kappa_- - \kappa_+) / 2\> , &(gmixsol)\cr
&\kappa_\pm = \left[ 1+ {g_{\ssc U}^2\over 8 \pi^2}
(9\pm\sqrt{6}\, )\; (t_U - t) \right ]^{-1/2} \> . &(whocares)
\cr }
$$
(The first equality in each of \(grsol) and \(gmixsol) is due to a
coincidental symmetry of the RG equations in this model.)
On this ``unification trajectory", the mixed couplings $\gBR$
and $\gRB$ remain fairly small ($<.04$).

Since the apparent unification [\cite{unification}]
of gauge couplings observed at LEP can be maintained in
this model, it is sensible to explore features of the low-energy theory
which follow from unified supergravity-inspired [\cite{supergravity}]
boundary conditions.
These boundary conditions include the supposition that at some scale
$M_U \ge 2 \times 10^{16}$ GeV the scalars in the theory have a common
soft supersymmetry breaking (mass)$^2$ (denoted $m_0^2$) and there is
a common mass $m_{1/2}$ for each gaugino. One can then integrate
the RG equations from $M_U$ down to the electroweak scale, and study the
resulting low-energy theory. Here we will restrict ourselves to some brief
comments regarding the impact of the extension of the MSSM of section 2,
using the MSSM (with no new fields below $M_U$) as a template.

It is possible to show that the well-known gaugino mass unification
prediction
$$
(M_3/g_3^2) = (M_2/g_2^2) = (M_1/g_Y^2) = (m_{1/2}/g^2_U)
\eqno(gauginomassun)
$$
at low energies is precisely maintained by the one-loop RG equations
of this model along the unification trajectory, provided
that the gaugino masses are unmixed at the scale $t_U$ in the $R$,
$B-L$ basis.
[There {\it is} mixing induced among the gaugino mass parameters in
the $R$, $B-L$ basis by RG running, yet the surviving $U(1)_Y$
gaugino mass parameter does satisfy \(gauginomassun).]
Therefore the predictions for chargino, neutralino, and
gluino masses are essentially unaffected in the model of section 2,
compared to the MSSM as a template. The condition \(gauginomassun)
is modified by small two-loop corrections [\cite{MV},\cite{Yamada}],
of course.

The predictions for masses of squarks and sleptons are affected in two
ways. First, one has $D$-term contributions to scalar masses
[\cite{Drees}-\cite{KM}]
due to the spontaneous breaking of the $U(1)$ symmetry.
In the model of section 2, one  finds that each scalar $\phi_i$ obtains
a contribution to its (mass)$^2$ of
$$
\eqalignno{
\Delta m_i^2 = ({\overline m}^2 - m^2) \Bigl [
{3\over 20} X_i +{3\over 10} Y_i (\gR^2 &- \gB^2  + \gRB^2-\gBR^2
&(dterms)\cr
 & - \sqrt{1/6} \; [\gR \gBR +\gB \gRB])/g^2_X
\Bigr ]
\cr }
$$
where $X_i = {4\over 3} Y_i - {5\over 3} (B-L)_i$
and ${\overline m}^2 - m^2$ is evaluated at the scale $M_I$.
For the unification trajectory,
the term proportional to $Y_i$ is non-vanishing but small,
however it is important to keep
in mind that it need not be so with more general boundary conditions
on the gauge couplings.
The corrections \(dterms) should be added to the scalar masses at the
intermediate scale $M_I$ and must be renormalized down to the electroweak
scale; this turns out [\cite{KM}] to not affect
the contributions proportional
to $X_i$, while inducing a quite small change in the term proportional
to $Y_i$.

The presence of the additional $U(1)$ gauge interactions above $M_I$
also makes a contribution to scalar masses, because of terms
in the RG equations due to gaugino loops. Evaluating these contributions
for the unification trajectory of the RG equations (and taking into
account all mixing effects) one finds the following approximate results
for the slepton masses at the electroweak scale:
$$
\eqalignno{
m^2_{\tilde e_R} & = [m_0^2 + .15 m_{1/2}^2 - \sin^2\theta_W m_Z^2 \cos
2\beta]
- {1\over 20} ({\overline m}^2 - m^2) +
.015 m_{1/2}^2 &(mer)\cr
m^2_{\tilde e_L} & = [m_0^2 + .52 m_{1/2}^2 + (\sin^2\theta_W -\half )
m_Z^2 \cos 2\beta  ]
+{3\over 20} ({\overline m}^2 - m^2) +
.045 m_{1/2}^2 &(mel)\cr
m^2_{\tilde \nu} & = [m_0^2 + .52 m_{1/2}^2 +
\half  m_Z^2 \cos 2\beta ]
+{3\over 20} ({\overline m}^2 - m^2) +
.045 m_{1/2}^2 &(mel)\cr }
$$
In each case the first set of terms in square brackets is the result
for the template (MSSM) model.
Next is the $D$-term associated with breaking of the $U(1)$ gauge group
(neglecting the small contribution proportional to $Y$, and with
${\overline m}^2 - m^2$ evaluated at $M_I$), and the last term
is the additional contribution from $U(1)$ gaugino loops above $M_I$.
We have used here representative values $M_U = 2\times 10^{16}$ GeV
and $M_I = 10^{10}$ GeV.
Similar equations can be written down for the
squarks, although the relative effect is much
larger for the sleptons. As long as we are assuming scalar mass unification,
there is good motivation for the expectation
that ${\overline m}^2 > m^2$, since $m^2$ receives a negative RG contribution
proportional to $|y_S|^2$ [c.f.~eq.\(seesaw)]. Therefore, the change
(compared to the MSSM) in the difference between charged slepton masses,
$$
\Delta(m^2_{\tilde e_L} - m^2_{\tilde e_R}) \approx
({\overline m}^2 - m^2)/5 \; + \; .03 m_{1/2}^2
\eqno(mseldiff)
$$
is expected to be positive. This reinforces the expectation in the MSSM
that $\tilde e_L$ should be heavier than $\tilde e_R$; depending on the
relative magnitudes of $m^2_0$, $m^2_{1/2}$, and ${\overline m}^2 - m^2$,
the difference \(mseldiff) could even be dramatic.

\subhead{4. An extension of the  minimal model}
\taghead{4.}

The model described in the section 2 is the simplest extension
of the MSSM with gauged $B-L$ breaking to matter parity at an intermediate
scale. Its other successful features include a natural solution
to the $\mu$ problem and a potentially successful theory of neutrino masses.
It is interesting to consider extensions of the minimal model with
an enlarged gauge symmetry at the symmetry breaking scale. Note, however,
that larger gauge groups seem to be somewhat disfavored for the following
reasons. In order to have a viable seesaw mechanism,
the order parameter field $S$ ought to be in a symmetric product
of the conjugate of the representation which contains $\nu$.
If the gauge group is extended to contain additional non-abelian factors,
such a symmetric product representation will generally be large,
containing fields which are not neutral under the Standard Model gauge group.
This jeopardizes asymptotic freedom of the gauge couplings,
and in any case forces us to view the apparent unification of gauge couplings
observed at LEP as merely a perverse accident. Furthermore, symmetric
product representations are quite difficult to obtain in string models.
Even worse, the positive gaugino-loop radiative corrections to soft
scalar masses are proportional to the quadratic Casimir invariant,
and thus tend to be large for symmetric product representations
containing $S$ and $\overline S$.
This effect seems to strongly disfavor VEVs
for such scalars, although sufficiently large term(s) analogous to the
$A$ term in \(vsoft) might overcome it.

Therefore, we will consider here only
the next-to-smallest gauge group containing gauged $U(1)_{B-L}$, namely
$SU(3)_C \times SU(2)_L \times SU(2)_R \times U(1)_{B-L}$.
Under this gauge group, the chiral superfields
transform as\footnote{$^\dagger$}{This is not a left-right symmetric
model, given our (minimal) choice of particle content. Similar
models are considered in [\cite{KM2}], but our
treatment will be somewhat different.}
$$
\eqalign{
&Q \sim ({\bf 3}, {\bf 2}, {\bf 1}, \frac1/3 ) \cr
&u,d \sim (\overline{\bf 3}, {\bf 1}, {\bf 2}, -\frac1/3 ) \cr
&S \sim ({\bf 1}, {\bf 1}, {\bf 3}, -2 ) \cr}\qquad\>\>\>
\eqalign{
&L \sim ({\bf 1}, {\bf 2}, {\bf 1}, -1 ) \cr
&e, \nu \sim ({\bf 1}, {\bf 1}, {\bf 2}, 1 ) \cr
&{\overline S} \sim ({\bf 1}, {\bf 1}, {\bf 3}, 2 )\> . \cr}\qquad\>\>\>
\eqalign{
&H_u,H_d \sim ({\bf 1}, {\bf 2}, {\bf 2}, 0 )\cr
&\phantom{nothing}\cr
&\phantom{zilch}\cr}
$$
The lowest-order non-renormalizable superpotential for the $S,\overline S$
degrees of freedom contains two independent terms in general:
$$
W = {\lambda_1\over 2 M} {\rm Tr}[S {\overline S}]^2 +
    {\lambda_2\over 4 M} {\rm Tr}[S^2]\> {\rm Tr}[{\overline S}^2]\> .
\eqno(superpotential2)
$$
Here we use a notation in which $SU(2)_R$ triplets are given by
traceless
$2\times 2$ matrices, explicitly
$$
S = \pmatrix{ S^{-}/\sqrt{2} & S^0 \cr
              S^{--} & -S^-/\sqrt{2} \cr }\> ;
\qquad\qquad
{\overline S} = \pmatrix{ {\overline S}^{+}/\sqrt{2} & {\overline S}^{++} \cr
              {\overline S}^{0} & -{\overline S}^+/\sqrt{2} \cr }
$$
with the superscripts indicating the electric charge. The soft breaking terms
are given by
$$
V_{\rm soft} = m^2 {\rm Tr}[S^\dagger S]
+ {\overline m}^2 {\rm Tr}[ {\overline S}^\dagger {\overline S}] -
\left ( {A_1\over 2 M} {\rm Tr}[S {\overline S}]^2 +
    {A_2\over 4 M} {\rm Tr}[S^2]\> {\rm Tr}[{\overline S}^2]
 + c.c. \right )\> .
\eqno(soft2)
$$
By a suitable phase rotation, we take the parameter
$A_1$ to be real and positive, while the phases of $\lambda_1,\lambda_2,
A_2$ are arbitrary.
There is a possible minimum of the full scalar potential
for the neutral scalar components of $S$ and $\overline S$
with VEVs in a nearly $D$-flat direction:
$$
\langle S^0 \rangle^2 \approx
\langle {\overline S}^0 \rangle^2 \approx
{M \over 6 |\lambda_1 |^2}
\left ( A_1 + \sqrt{ A_1^2 - 6 |\lambda_1 |^2 (m^2 +
{\overline m}^2 ) } \right ) \> ,
\eqno(vevS2)
$$
with the deviation from $D$-flatness given by
$$
\langle S^0 \rangle^2 -
\langle {\overline S^0} \rangle^2 \approx
({\overline m}^2 - m^2)/(3 g_{B-L}^2 + 2 g_R^2) \> .
\eqno(diffvev2)
$$
This minimum is stable against local perturbations provided that
$$
\eqalignno{
&A_1^2 - 6 |\lambda_1 |^2 (m^2 +
{\overline m}^2 )> 0\> ,  &(stability1)\cr
& |\lambda_1 + \lambda_2 |^2 s^2 + {m^2 + {\overline m}^2 \over 2} >
\Bigl [s^2 |A_1 + A_2 - 2\lambda_1^*(\lambda_1 + \lambda_2) s|^2
 + 4 g_R^4 \Delta^4 \Bigr ]^{1/2}
\> .
&(stability2) \cr }
$$
where $s = \langle S^0 \rangle^2/M$ and $\Delta^2 =\langle S^0 \rangle^2
- \langle {\overline S}^0 \rangle^2 $ define two convenient
parameters of order $m_W$.
These stability conditions are satisfied in a non-vanishing region
of the parameter space. In a smaller, but still non-vanishing, region
of parameter space this is also a global minimum of the potential.
(However, it is not clearly relevant to require that the desired
minimum be global, since the lifetime of the false vacuum might be many
orders of magnitude longer than the age of the universe.)
The VEVs break the gauge symmetry according to
$$
SU(2)_R \times U(1)_{B-L} \rightarrow
U(1)_Y \times ({\rm matter} \> \>{\rm parity})\> .
$$
A viable spectrum of neutrino masses can arise just as before
via the seesaw mechanism, by taking the obvious extension of
\(seesaw).
The $\mu$ term of the MSSM can also
arise in a way exactly analogous to
that discussed in the model of the previous section, from
a non-renormalizable superpotential term proportional to
$H_u H_d {\rm Tr}[S {\overline S}]$.

The resulting spectrum contains intermediate-scale states
consisting of a pair of charge $\pm 1$ massive vector
supermultiplets with mass ${\sqrt 2}\, g_R \langle S \rangle$ and a
neutral massive vector supermultiplet of mass
$\sqrt{6g_{B-L}^2+ 4 g_R^2}\, \langle S \rangle $. The remaining uneaten
components of $S$ and $\overline S$ obtain electroweak-scale masses,
and consist of the pair of charge $\pm 2$ chiral
supermultiplets $S^{--}$, ${\overline S}^{++}$
and one neutral chiral supermultiplet $\Phi$. The components of $\Phi$
get masses just given by eqs.~\(psimass)-\(massab)
with $\lambda \rightarrow \lambda_1$
and $A \rightarrow A_1$; and all of the same comments apply as before.
The fermionic components of $S^{--}$ and ${\overline S}^{++}$
obtain masses $|\lambda_1 + \lambda_2| s$,
while their light scalar partners have squared masses
$$
|\lambda_1 + \lambda_2 |^2 s^2 + {m^2 + {\overline m}^2 \over 2} \pm
\Bigl [s^2 |A_1 + A_2 - 2\lambda_1^*(\lambda_1 + \lambda_2) s|^2
 + 4 g_R^4 \Delta^4 \Bigr ]^{1/2}
\> .
\eqno(uglymasses)
$$
These masses suffer renormalization between the scale of
spontaneous symmetry breaking and the electroweak scale,
since $S^{--}$, ${\overline S}^{++}$
are charged under the MSSM gauge group.

The most
striking prediction of this model is therefore the presence of an exotic
vector-like pair of chiral supermultiplets of electric charge $\pm 2$
which may well be accessible to future collider experiments.
These particles
will have unsuppressed
two-body decays into pairs of like-sign leptons, because
of the superpotential interaction
$
W \supset y_S eeS^{--}
$
which derives from the analog of \(seesaw). This should yield a striking
experimental signature.
This feature is shared
by left-right symmetric models with symmetry breaking
near the electroweak scale [\cite{HM}]. In the present case, the lightness of
these
exotic states is due to the lack of renormalizable mass couplings in
the underlying superpotential. One can check that the presence of these
exotic states does not cause the gauge couplings to blow up below the
Planck scale; however, they do completely modify
the running. In this model, gauge coupling
unification in the usual sense would require additional fields not
considered here, and unlike in the model of section 2,
the LEP observation of apparent unification would
have to be viewed as entirely accidental.

\subhead{5. Conclusion}
\taghead{5.}

In this paper we have analyzed what might be called the minimal
supersymmetric standard model with automatic $R$-parity conservation.
We do not include  renormalizable tree-level mass terms in the
superpotential. Instead, gauge
symmetry breaking arises because of the interplay
between non-renormalizable interactions and soft supersymmetry-breaking
interactions. We then found that it is important to take into account
dimensionless but ``soft" supersymmetry-breaking couplings in this
analysis, which can play a crucial role in the spontaneous symmetry
breaking; indeed, it may not be able to understand the symmetry breaking
mechanism without them. These models have several
attractive features. First, the $\mu$ term of the MSSM is naturally
generated by a mechanism familiar from [\cite{KN}]. Second,
the masses of neutrinos are determined by an intermediate-scale
seesaw mechanism and so may be phenomenologically interesting.
The minimal version of the model in section 2 also has the nice
property that the apparent unification of gauge couplings
observed at LEP can still be considered non-accidental.
In this model we found that, assuming supergravity-inspired
boundary conditions on the soft terms, a discernible imprint
may be left on the spectrum of MSSM sparticles. In particular,
the masses of the left-handed sleptons are further increased over those
of the right-handed sleptons.

One of the interesting consequences of this class of models
is the existence of a supermultiplet ($\Phi$) of neutral particles with
electroweak scale masses and only very weak couplings to MSSM particles.
The largest couplings of these particles to MSSM states are
suppressed by at least $(m_W/M)^{1/2}$, and arise from the same
non-renormalizable interaction which induces the $\mu$ term of the MSSM.
These particles should therefore have relatively long (perhaps
microsecond or nanosecond) lifetimes. While this may provide for
interesting cosmological consequences, the weak couplings of
these particles means that they cannot play a role in collider
experiments. In the next-to-minimal model, we found that the symmetry
breaking mechanism also predicts a pair of exotic chiral supermultiplets
of particles with electric charge $\pm 2$ and electroweak-scale masses.

It is possible that $R$-parity conservation
cannot be ``explained", but should simply be taken as a law of nature.
It is also possible that an explanation exists, but lies only on
the far side of the Planck or string scale. However, it is gratifying
that one can construct field theory
models which are consistent with all known observations, and
in which $R$-parity conservation has its origin in terms of a deeper
gauge principle.
This may be taken as one of many clues to the nature of
physics at very high energies.

I am grateful to James D. Wells for many helpful discussions.
This work was supported in part by the U.S. Department of Energy.

\hyphenation{N-i-l-l-e-s}
\hyphenation{V-a-l-l-e}
\hyphenation{M-o-r-a-w-i-t-z}
\references
\oneandtwofifthsspace

\refis{unification} P.~Langacker, in Proceedings of the
PASCOS90 Symposium, Eds.~P.~Nath
and S.~Reucroft, (World Scientific, Singapore 1990)
J.~Ellis, S.~Kelley, and D.~Nanopoulos, \pl B260, 131, 1991;
U.~Amaldi, W.~de Boer, and H.~Furstenau, \pl B260, 447, 1991;
P.~Langacker and M.~Luo, \pr D44, 817, 1991.

\refis{alternative}
M.~C.~Bento, L.~Hall and  G.~G.~Ross, \np B292, 400, 1987.

\refis{baryonparity} L.~E.~Ib\'a\~nez and G.~G.~Ross, \np B368, 3, 1992.

\refis{discreteanomalies} L.~E.~Ib\'a\~nez and G.~G.~Ross, \pl B260, 291, 1991.

\refis{Mohapatra} R.N.~Mohapatra, \pr D34, 3457, 1986.

\refis{seesaw} M.~Gell-Mann, P.~Ramond, and R.~Slansky, in Sanibel Talk,
CALT-68-709, Feb 1979, and in {\it Supergravity},
(North Holland, Amsterdam, 1979; T.~Yanagida, in {\it Proc. of the
Workshop on
Unified Theories and Baryon Number in the Universe}, Tsukuba, Japan, 1979,
edited by A. Sawada and A. Sugamoto (KEK Report No. 79-18, Tsukuba, 1979).

\refis{Drees} M.~Drees, \pl B181, 279, 1986.

\refis{HK} J.~S.~Hagelin and S.~Kelley, \np B342, 95, 1990.

\refis{FHKN} A.~E.~Faraggi, J.~S.~Hagelin, S.~Kelley, and
D.~V.~Nanopoulos, \pr D45, 3272, 1992.

\refis{KT} Y.~Kawamura and M.~Tanaka,
\journal Prog. Theor. Phys., 91, 949, 1994;
Y.~Kawamura, H.~Murayama, M.~Yamaguchi, \pl B324, 52, 1994;
\pr D51, 1337, 1995.

\refis{KM} C.~Kolda and S.~P.~Martin, ``Low-energy supersymmetry
with D-term contributions to scalar masses'', preprint
hep-ph/9503445, Phys. Rev. D, to appear.

\refis{KW} L.~Krauss and  F.~Wilczek, \prl 62, 1221, 1989.

\refis{sscfgrp} S.~P.~Martin \pr  D46, 2769, 1992.

\refis{violation1}
F.~Zwirner, \pl B132, 103, 1983;
L.~J.~Hall and M.~Suzuki, {\it Nucl. Phys.} {\bf B231} (1984), 419;
I.~H.~Lee, \np B246, 120, 1984;
S.~Dawson, \np B261, 297, 1985;
R.~Barbieri and A.~Masiero, \np B267, 679, 1986;
S.~Dimopolous, L.~J.~Hall, \pl B207, 210, 1987;
I.~Hinchliffe and T.~Kaeding, \pr D47, 279, 1993;
J.~L.~Goity and M.~Sher, \pl B346, 69, 1995;
C.~E.~Carlson, P.~Roy, and M.~Sher, \pl B357, 99, 1995.

\refis{violation4} J.~Butterworth and H.~Dreiner, \np B397, 3, 1993;
H.~Dreiner and P.~Morawitz, \np B428, 31, 1994.

\refis{violation3}
S.~Dimopolous et. al., \pr D41, 2099, 1990;
H.~Dreiner and  G.~G.~Ross, \np B365, 597, 1991;
V.~Barger, M.~S.~Berger, P.~Ohmann, and R.~J.~N.~Phillips,
\pr D50, 4299, 1994;
H.~Baer, C.~Kao, and X.~Tata, \pr D51, 2180, 1995.

\refis{violation2}
V.~Barger, G.~F.~Giudice, and T.~Han, \pr D40, 2987, 1989;
R.~Barbieri, D.~E.~Brahm, L.~J.~Hall and S.~D.~H.~Hsu, \pl B238, 86, 1990;
R.~M.~Godbole, P.~Roy, and X.~Tata, \np B401, 67, 1993.

\refis{rparity} G.~Farrar, P.~Fayet, \pl B76, 575, 1978.

\refis{rparity2} S.~Dimopolous and H.~Georgi, \np B193, 150, 1981;
S.~Weinberg, \pr D26, 287, 1982;
N.~Sakai and T.~Yanagida, \np B197, 533, 1982.

\refis{FIQ}  A.~Font, L.~E.~Ib\'a\~nez and F.~Quevedo, \pl B228, 79, 1989.

\refis{more}  L.~E.~Ib\'a\~nez, \npb 398, 301, 1993.

\refis{BD} T.~Banks and M.~Dine, \pr D45, 1424, 1995.

\refis{MSY} H.~Murayama, H.~Suzuki, and T.~Yanagida,
\pl B291, 418, 1992.

\refis{spon} C.~S.~Aulakh and R.~N.~Mohapatra,
\pl B119, 136, 1982;
J.~Ellis, G.~Gelmini, C.~Jarlskog, G.~G.~Ross, and J.~W.~F.~Valle,
\pl B150, 142, 1985;
G.~G.~Ross and J.~W.~F.~Valle,
\pl B151, 375, 1985;
B.~Gato, J.~Leon, J.~Perez-Mercader, and M.~Quiros,
\np B260, 203, 1985;
A.~Masiero and J.~W.~F.~Valle,
\pl B251, 273, 1990.

\refis{KN} J.~E.~Kim and H.~P.~Nilles, \pl B138, 150, 1984.

\refis{MV} S.~P.~Martin and M.~T.~Vaughn, \pl B318, 331, 1993.

\refis{Yamada} Y.~Yamada, \pl B316, 109, 1993, \prl 72, 25, 1994.

\refis{CM} D.~J.~Casta\~no and S.~P.~Martin, \pl B340, 67, 1994.

\refis{KM2} R.~Kuchimanchi and R.~N.~Mohapatra, \pr D48, 4352, 1993;
\prl 75, 3989, 1995; R.~Kuchimanchi, preprint hep-ph/9511376;
R.~N.~Mohapatra and A.~Rasin, preprint hep-ph/9511391.

\refis{HM} See for example
K.~Huitu, J.~Maalampi, and M.~Raidal, \np B420, 449, 1994;
K.~Huitu and J.~Maalampi, \pl B344, 217, 1995, and references therein.

\refis{mixing} See F.~del Aguila, G.~D.~Coughlan, and M.~Quir\'os,
\np B307, 633, 1988;
F.~del Aguila, M.~Masip, and M.~P\'erez-Victoria,
preprint hep-ph/9507455, and references therein.

\refis{Langacker} See for example P.~Langacker, ``Neutrino Physics", talk
presented at Beyond the Standard Model IV, Lake Tahoe, December 1994, and
references therein.

\refis{supergravity}
A.~Chamseddine, R.~Arnowitt and P.~Nath, \prl 49, 970, 1982;
H.~P.~Nilles, \pl 115B, 193, 1982;
L.~E.~Ib\'a\~nez, \journal Phys.~Lett., 118B, 73, 1982;
R.~Barbieri, S.~Ferrara, and C.~Savoy, \pl 119B, 343, 1982;
L. Hall, J. Lykken and S. Weinberg, \pr D27, 2359, 1983;
P. Nath, R. Arnowitt and A. H. Chamseddine, \np B227, 121, 1983.

\endreferences
\endit\end